\documentclass[twocolumn,aps,showpacs,preprintnumbers,superscriptaddress]{revtex4}
\usepackage{graphicx}
\usepackage{bm}
\usepackage{amsmath, amsthm, amssymb}
\usepackage{color}

\begin{document}
\title{Optical symmetric pushing, uni-/bi-directional null, and pulling-pushing flipped forces in one dimensional PT-symmetric photonics}
\author{Jeng Yi Lee}
\affiliation{Department of Opto-Electronic Engineering, National Dong Hwa University, Hualien 974301, Taiwan }
\author{Pai-Yen Chen}
\affiliation{Department of Electrical and Computer Engineering, University of Illinois at Chicago, Chicago, IL 60661,
USA}

\date{\today}

\begin{abstract} 
We discuss the optical forces exerted on parity-time (PT) symmetric heterostructures under normal incidence of a single and two counter-propagating plane waves.
The underlying strategy is through generalized parametric space, stemming from consideration of PT-symmetry condition and Lorentz reciprocity theorem.
In such a generalized parametric space, we are able to not only exhaustively indicate various PT phases and extraordinary wave phenomena, but also deduce the directionality and magnitudes of optical forces.
We find that when the system is illuminated by a normally incident wave, it can exhibit the symmetric pushing effect in the exact symmetry phase, uni-directional null (UNF) and  bi-directional null force (BNF) at the exceptional point (EP), and pulling-pushing flipped forces in the broken symmetry phase, with BNF found at the pushing-pulling turning point.
In two counter-propagating plane waves interference, the magnitudes as well as the directionality of resultant optical force can be tuned by relative phase of incident waves.
More interestingly, we observe that there has a null force independent of the relative phase occurred in a specific region of broken symmetry phase and exceptional point.
In addition, we offer several PT-symmetric heterostructures to support our findings.
Our results may be beneficial for applications in PT optomechanics and force rectifiers.  
\end{abstract}
\pacs{ }

\maketitle

\section{Introduction}
Optical pulling force, i.e., negative optical pressure, is a counter-intuitive phenomenon, due to that photons deliver linear momentum to matters.
A plane wave upon passive slabs and particles is unable to generate optical pulling force \cite{book1,book2,nobel1,particle1}.
Several architectures have demonstrated to have this optical pulling force by engineering structured beams or gain media embedded \cite{beam1,beam2,beam3,plane1,plane2,plane3,beam5,beam6,gain1,gain2,gain3}.
In the former case, many efforts have been made on nondiffracting Bessel beam \cite{beam1,beam2}, tractor beams \cite{beam3}, multiple plane waves interference \cite{plane1,plane2,plane3},  solenoid beam \cite{beam5}, and  temporal decaying pulse excitation \cite{beam6}.
For the latter one, optical pulling force exerted on gain slab and particle can  emerge in vicinity of Fabry-Perot and Fano resonances, respectively \cite{gain1,gain3}.
We note that in complex frequency excitation, optical pulling force could appear when temporal decay rate of stored energy in cavity is slower than that of impinging temporal decay wave, equivalent to virtual gain \cite{beam6}.
In principle, achievement of optical pulling force on slabs (particles) relies on suppression of a backward wave, i.e., reduced reflectance (backward-scattering), and enhancement of a forward wave, i.e., enhanced transmittance (forward-scattering).

 When a scattering system is made of gain and loss media, the corresponding Hamiltonian is non-Hermitian. 
 With a balance of gain and loss indexes of refraction in spatial placement,  i.e., $n(\vec{r})=n^{*}(-\vec{r})$, the whole system has a parity-time symmetry (space-time reflection symmetry, PT in shorthand) \cite{pt1}.
For an importing wave transversely propagating to a one dimensional PT symmetric system, its optical responses can be classified into symmetric phase and broken symmetric phase, while there has an exceptional point in between  \cite{pt1,pt2,pt3}.
 PT-symmetric systems have extraordinary functionalities, such as  coherent perfect absorption and lasing (CPAL) \cite{pt2,pt4}, unidirectional invisibilities \cite{pt5}, power oscillation \cite{pt6}, Bloch oscillation \cite{pt7}, 
negative refraction \cite{pt8}, ulta-sensitive sensing \cite{pt9,pt10},  optical isolation \cite{pt11}, and conical diffraction \cite{pt12}.
Moreover, PT-symmetric systems can support asymmetric reflectances and beyond-unity transmittance, that one can anticipate the emergence of asymmetric pushing and  pulling forces \cite{ptforce1,ptforce2}.
Quite counter-intuitively, we recently find that when PT-symmetric systems operating at a specific regime of symmetric phase and exceptional point, one can have non-distinguishable complex reflection coefficients from two opposite incidences, even although PT-symmetric system can not satisfy parity symmetry alone\cite{pt13}.
Hence, instead of concerning specific PT-symmetric systems by a plane wave illumination, understanding limitation of optical force in any PT-symmetric systems as  well as two counter-propagating plane waves incidence is of importance.

 In this work, we link the optical force with PT phases.
 Our discussion framework, for the first time, is  beyond any specific PT-system configurations, material parameters, and operating frequency.
 The underlying means is by generalized parametric space, originating from consideration of  PT symmetry condition and Lorentz reciprocity theorem.
In symmetric phase and a plane wave from two opposite incident directions, all optical forces are pushing.
PT symmetric systems operating at parity symmetry of reflection (PSR) can have equal magnitudes of the optical pushing force by two individually opposite illuminations, i.e., symmetric pushing force, while others have asymmetric pushing force.
In the broken symmetric phase, we observe that there has a pulling-pushing flipped force, i.e., pushing from one incident direction but pulling from another incident direction.
We also find that there has a bi-directional null force (BNF) within a regime of broken symmetric phase and exceptional point.
In an exceptional point, the optical force can be pushing or null, depending on incident direction.
For the latter case, this is due to the absence of reflection  as well as unity transmittance.
In two opposite propagating plane waves illumination, we find that directionality and magnitudes of optical forces can be altered by tuning relative phase of waves.
However, quite interestingly, we discover there has a null force independent of the relative phases occurred in a specific regime of broken symmetric phase and exceptional point.
In addition, we provide several PT-symmetric systems made of different system configurations, material parameters, and operating frequencies to support our findings.
We believe our work can provide benefit in the design of PT-symmetric optomechanics.

\begin{figure*}[t]
\centering
\includegraphics[width=17cm]{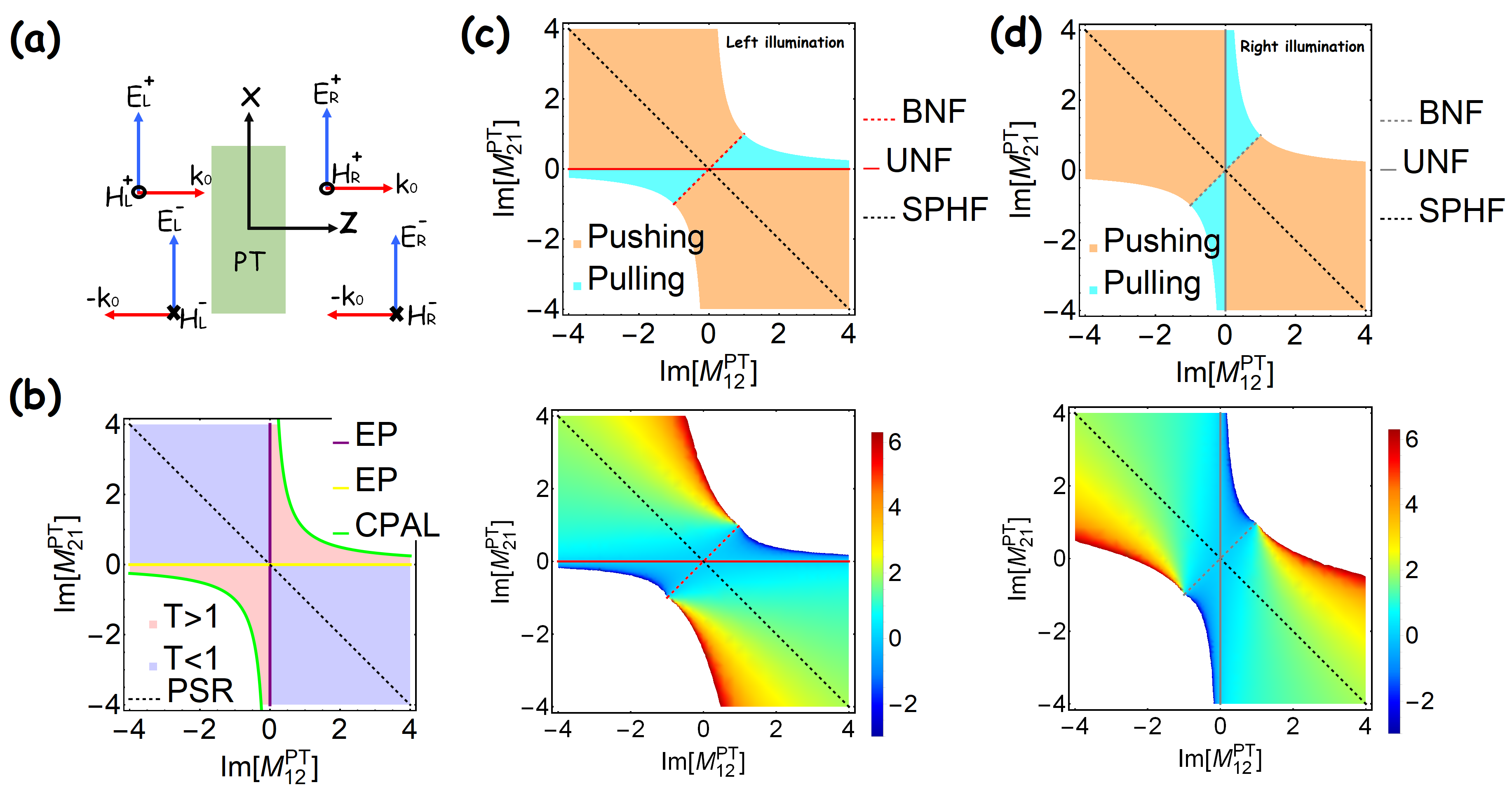}
\caption{(Color online) (a) Schematic of two independent x-polarized electromagnetic plane waves  $E_L^{+}$ and $E_R^{-}$ normal incident to an extended infinite planar slab with PT symmetry. The corresponding complex amplitudes of outgoing plane waves are $E_L^{-}$ and $E_R^{+}$. Through concerning PT symmetry condition and Lorentz reciprocity theorem into the transfer matrix, we propose a generalized parametric space in (b). White region presents forbidden parameters:  no PT systems can have this result.
Light-purple region represents the PT symmetric phase accompanied with $T<1$, while Pink-color region denotes PT broken symmetric phase with $T>1$.
A black dashing line corresponds to the parity-symmetry of reflectances (PSR), i.e., $r_R=r_L$.
Between the PT symmetric and broken symmetric phases, there is exceptional point, depicted by yellow and dark-purple lines.
The exceptional point supports uni-directional reflectionless and unity transmittance $T=1$.
However, there can support bi-directional reflectionless and unity transmittance at $Im[M_{12}]=Im[M_{21}]=0$, corresponding to the cross point by two exceptional points.
 Within the framework of this generalized parametric space, with Eqs.(8) and (9), we reveal the directionality and magnitudes of the optical forces $\vec{P}_L$ and $\vec{P}_R$ illuminated by a plane wave from the left and right leads in (c) and (d), respectively.
Dark-orange color means the optical pushing force, while water-blue color means the optical pulling force.
We highlight the optical pushing force with equal magnitudes $\vert\vec{P}_L\vert=\vert\vec{P}_R\vert$ by a black dashing line, i.e., symmetric pushing force (SPHF).
We also depict a red and gray solid lines located in the exceptional point in (c) and (d) for uni-directional null force (UNF), accompanied with left-reflectionless and right-reflectionless as well as unity transmittance, respectively.
Interestingly, there has a bi-directional null force (BNF), marked by a red and gray dashing lines, located in the specific regime of broken symmetric phase and exceptional point.
} 
\end{figure*}  

\section{Theoretical framework: depiction of directionality and magnitudes of optical forces on arbitrary PT-symmetric systems by means of generalized parametric space}
We consider two independent x-polarized electromagnetic plane waves propagating toward  $\pm z$ directions to a planar slab system with PT symmetry, as shown in Fig. 1 (a).
Here $E_L^{+}$ ($E_L^{-}$) and $E_R^{+}$ ($E_R^{-}$)  represent complex amplitudes for right-propagating (left-propagating) plane waves $e^{ik_0 z}$ ($e^{-ik_0z}$) in the left and right leads, respectively.
Outside PT symmetric system, environment is vacuum with wavenumber $k_0$, permittivity $\epsilon_0$, and permeability $\mu_0$.
Throughout this work, time evolution of all electromagnetic plane wave is $e^{-i\omega t}$, here $\omega$ is angular frequency.

Description of these plane waves interacted with an infinitely extended slab, by transfer matrix formalism, we have
\begin{equation}
\begin{split}
\begin{bmatrix}
E_R^{+}\\
E_R^{-}
\end{bmatrix}&=
M\begin{bmatrix}
E_L^{+}\\
E_L^{-}\\
\end{bmatrix}.\\
\end{split}
\end{equation}
Then, by introducing PT symmetry, we know the corresponding transfer matrix has $(M^{PT})^{-1}=(M^{PT})^{*}$ \cite{pt3,pt4}. 
Further, applying Lorentz reciprocity theorem, we get $Det[M^{PT}]=1$.
So the elements of arbitrary PT symmetric transfer matrices can be expressed as
\begin{equation}\label{parametric}
\begin{split}
M^{PT}_{12(21)}&=iIm[M^{PT}_{12(21)}],\\
M^{PT}_{11}&=M^{PT*}_{22}=\sqrt{1-Im[M^{PT}_{12}]Im[M^{PT}_{21}]}e^{i\phi}
\end{split}
\end{equation}
where $\phi$ is transmission phase.
As a result, description of any PT symmetric transfer matrix needs only three real parameters: $Im[M^{PT}_{12}]$, $Im[M^{PT}_{21}]$, and $\phi$, with
\begin{equation}
\begin{split}
Im[M^{PT}_{12}]&Im[M^{PT}_{21}]\leq 1,\\
\phi & \in[-\pi,\pi].
\end{split}
\end{equation}

The corresponding PT symmetric scattering matrix $S^{PT}$ \cite{pt3,pt13}, linking incoming waves $[E_L^{+},E_R^{-}]^T$ with outgoing waves $[E_R^{+},E_L^{-}]^T$, in terms of the PT transfer matrix formalism, is 
\begin{equation}\label{scattering}
\begin{split}
\begin{bmatrix}
E_R^{+}\\
E_L^{-}
\end{bmatrix}&= S^{PT}\begin{bmatrix}
E_L^{+}\\
E_R^{-}\\
\end{bmatrix}=
\begin{bmatrix}
t & r_R\\
r_L & t
\end{bmatrix}
\begin{bmatrix}
E_L^{+}\\
E_R^{-}\\
\end{bmatrix}\\
&=\begin{bmatrix}
\frac{e^{i\phi}}{\sqrt{1-Im[M^{PT}_{12}]Im[M^{PT}_{21}]}} & \frac{Im[M^{PT}_{12}]e^{i(\phi+\frac{\pi}{2})}}{\sqrt{1-Im[M^{PT}_{12}]Im[M^{PT}_{21}]}}\\
\frac{Im[M^{PT}_{21}]e^{i(\phi-\frac{\pi}{2})}}{\sqrt{1-Im[M^{PT}_{12}]Im[M^{PT}_{21}]}} & \frac{e^{i\phi}}{\sqrt{1-Im[M^{PT}_{12}]Im[M^{PT}_{21}]}}
\end{bmatrix}\begin{bmatrix}
E_L^{+}\\
E_R^{-}\\
\end{bmatrix}
\end{split}
\end{equation}
here $t$ is complex transmission coefficient, $r_L$ and $r_R$ are complex left and right reflection coefficients, respectively. 
Now, the corresponding scattering eigenvalues and scattering eigenvectors of $S^{PT}$ are $S^{PT}_{\pm}=\frac{e^{i\phi}}{\sqrt{1-Im[M^{PT}_{12}]Im[M^{PT}_{21}]}}(1\pm \sqrt{Im[M^{PT}_{12}]Im[M^{PT}_{21}]})$ and $\vert S^{PT}_{\pm}>=[\pm i\sqrt{\frac{Im[M^{PT}_{12}]}{Im[M^{PT}_{21}]}},1]^T$, respectively \cite{pt3,pt13}.

As discussion in Refs. \cite{pt3,pt13}, various anomalous scattering phenomena  can have any transmission phases. 
Moreover, these anomalous wave phenomena are crucially associated with two real parameters:$Im[M^{PT}_{12}]$ and $Im[M^{PT}_{21}]$.
In the following analysis, we would show  that these two real parameters $Im[M^{PT}_{12}]$ and $Im[M^{PT}_{21}]$, not transmission phase $\phi$, would be involved in the optical forces by a single and two counter-propagating plane waves excitation.
Hence we accordingly depict the parametric space as shown in Fig. 1 (b).
Colour region denotes allowable parameters  for any PT-symmetric systems, while white denotes forbidden parameters.

When a PT-symmetric system operating at $Im[M^{PT}_{12}]Im[M^{PT}_{21}]<0$, we see that the transmittance $T=\vert t \vert^2<1$ and the scattering eigenvalues become unimodulus and distinguishable, corresponding to symmetric phase.
We indicate this symmetric phase region depicted in the light-purple color of Fig. 1 (b).
On the other hand, when a PT symmetric system is in $0< Im[M^{PT}_{12}]Im[M^{PT}_{21}]\leq 1$, the transmittance follows $T=\vert t \vert^2>1$ and the scattering eigenvalues form a reciprocal modulus pair and also distinguishable, belonging to broken symmetric phase, as shown in the pink color of Fig. 1 (b).
In this situation, one scattering eigenvalue corresponds to amplifying, while another attenuation.
In particular, when $Im[M^{PT}_{12}]Im[M^{PT}_{21}]=1$, equivalent to $M^{PT}_{11}=0$ and $M^{PT}_{22}=0$, one scattering eigenvalue is zero while another is pole (singularity), so the system has a hybrid functionality of coherent perfect absorption and lasing (CPAL) \cite{pt2,pt4}, shown in the green line of Fig. 1 (b).
Exhibition of coherent perfect absorption, i.e., all incoming waves are totally absorbed, would rely on the proper incoming waves whose the corresponding scattering eigenvalue is zero, otherwise the PT-symmetric system operating in CPAL would display amplifying.
This dramatic change is highly sensitive to input waves, with a great application in design of ultra-sensitive sensing, \cite{pt9}.
When $Im[M^{PT}_{12}]=0$ or $Im[M^{PT}_{21}]=0$, the scattering eigenvalues become degenerate (non-distinguishable) and transmittance becomes unity $T=1$, also with $r_R=0$ or $r_L=0$, respectively, as shown in the dark-purple and 	yellow lines of Fig. 1 (b).
This situation is exceptional point (EP), with unidirectional reflectionless and $T=1$.
We note that  bi-reflectionless  is enable when $r_R=r_L=0$ ($Im[M_{12}^{PT}]=Im[M_{21}^{PT}]=0$), occurred in the cross point by the dark-purple and 	yellow lines.
As pointed by \cite{pt3,pt13}, PT-symmetric system operating at an exception point can not be regarded as rigorous invisibility due to non-null transmission phases.
We also find that there enables to have $r_R=r_L$, i.e., parity symmetry of reflectances (PSR), located in a specific regime of  symmetry phase and exceptional point, as marked by the black dashed line of Fig. 1 (b) \cite{pt13}.
This result is counter-intuitive, due to that PT-symmetric system violates parity symmetry alone.

Now, we turn to discuss the optical force exerted on PT-symmetric planar slab encountered by two counter-propagating plane waves impinging, as shown in Fig. 1 (a).
To calculate the optical force, we use
\begin{equation}
<\vec{F}>_T=\oint_{s}<\hat{T}>_T\cdot \hat{n}da
\end{equation}
where this is a surface integration over a planar slab, $<\hat{T}>_T$ is time-average Maxwell's stress tensor, and $\hat{n}$ is normal unit vector perpendicular to the slab surface \cite{book2}.
The  Maxwell's stress tensor reads
\begin{equation}\label{tensor}
\begin{split}
\hat{T}=\epsilon_0\vec{E}\otimes\vec{E}+\mu_0\vec{H}\otimes\vec{H}-\frac{1}{2}(\epsilon_0E^2+\mu_0H^2)\hat{I}\\
\end{split}
\end{equation}
where $\vec{E}$ and $\vec{H}$ are real electric and magnetic fields, $\otimes$ is dyadic product, and  $\hat{I}$ is second-order unit tensor.
One note is that the calculation of the optical force is a self-consistent problem, which $\vec{E}$ and $\vec{H}$ need to involve total incident and scattering fields.
We assume that the slab body is rigid, without deformation under electromagnetic fields interaction.
Accordingly, we have the time-average optical force per unit area, i.e., radiation pressure $P_{\sum}$,
 \begin{widetext}
\begin{equation}\label{twowave}
\begin{split}
P_{\sum}\hat{z}=\frac{<\vec{F}>_T}{A}=\frac{1}{2}\epsilon_0 \hat{z}\{\vert E_L^{+} \vert^2 (1-T+R_L)-\vert  E_R^{-} \vert^2(1-T+R_R)+2Re[ E_L^{+}  E_R^{-*}r_Lt^{*} ]-2Re[E_L^{+} E_R^{-*}r_R^{*}t]\}
\end{split}
\end{equation}
\end{widetext}
where $A$ is area of the slab.
The detailed derivation is placed at the Appendix.
In a single wave incidence, the first term (second term) of Eq.(\ref{twowave}) i.e., $\vert E_L^{+} \vert^2 (1-T+R_L)$ ($\vert E_R^{-}\vert^2(1-T+R_R)$) can be regarded as a sum of linear momentum change of photons by individual left illumination (right illumination): reflected photons would be responsible for pushing, while transmitted photons would be for pulling.
In addition, the phases of incident, reflected, and transmitted waves would not affect this optical force.
However, it is more interesting to see that when encountering two counter-propagating plane waves interferences, the participation of the third and fourth terms, i.e., ($2Re[ E_L^{+}  E_R^{-*}r_Lt^{*} ]$ and $2Re[E_L^{+} E_R^{-*}r_R^{*}t]$) indicates that the relative phase of impinging waves would dominantly decide the optical force.
Later we would provide a detailed analysis.

\section{Optical force by a single plane wave incidence}
We first discuss the optical force by one input electromagnetic plane wave from left illumination, $\vec{P}_L$, i.e., $E_L^{+}=E_0$ and $E_R^{-}=0$ , so we have
\begin{equation}
\begin{split}
\vec{P}_L&=\frac{1}{2}\epsilon_0E_0^2(1-T+R_L)\hat{z}\\
&=\frac{1}{2}\epsilon_0E_0^2\frac{Im[M_{21}^{PT}](Im[M_{21}^{PT}]-Im[M_{12}^{PT}])}{1-Im[M_{12}^{PT}]Im[M_{21}^{PT}]}\hat{z}
\end{split}
\end{equation}
here $E_0$ is real.
Then with above formula, based on the generalized parametric space, we plot this optical force $\vec{P}_L$  shown in Fig. 1 (c).
Throughout this work, to simplify, we have let $\frac{1}{2}\epsilon_0E_0^2=1$.
In the upper panel of Fig. 1 (c), we reveal the optical pushing force marked by dark-orange color and the optical pulling force marked by water-blue color.
In the bottom panel, we also depict the magnitudes of the optical force.
Now, it is straightforward to see that any PT symmetric systems operating at the symmetric phase would all have the optical pushing force.
However, in the PT broken symmetric phase, we find there have three situations in the optical forces:  pushing, pushing, and null forces ($\vec{P}_L=0$), in Fig. 1 (c).
For the latter case, marked by a dashed-red line, it is because $Im[M_{12}^{PT}]=Im[M_{21}^{PT}]$.
In the exceptional point, the optical force can be pushing or uni-directional null force (UNF), depending on the incident direction.
This UNF can appear in the absence of left reflectionless, i.e.,$r_L=0$.

Now, we turn to discuss the optical force by right illuminating, $\vec{P}_R$.
In this case,  $E_L^{+}=0$ and $E_R^{-}=E_0$,
then the corresponding formula for the optical force is
\begin{equation}
\begin{split}
\vec{P}_R&=-\frac{1}{2}\epsilon_0E_0^2(1-T+R_R) \hat{z}\\
&=-\frac{1}{2}\epsilon_0E_0^2 \frac{Im[M_{12}^{PT}](Im[M_{12}^{PT}]-Im[M_{21}^{PT})]}{1-Im[M_{12}^{PT}]Im[M_{21}^{PT}]}\hat{z}
\end{split}
\end{equation}
here the directionality of $\vec{P}_R$ toward $-\hat{z}$ denotes optical pushing force, while that toward $+\hat{z}$ denotes optical pulling force.
We also depict the directionality and magnitudes of $\vec{P}_R$ in Fig. 1 (d).
It clearly shows that in the symmetric phase, all optical force are pushing.
However, compared Fig. 1 (c) with Fig. 1 (d), we observe that this optical pushing force is asymmetry, i.e., unequal magnitudes of the optical forces $\vert\vec{P}_L\vert\neq\vert \vec{P}_R\vert$.
However, at PSR, the optical pushing forces become symmetry, i.e., $\vert\vec{P}_L\vert=\vert \vec{P}_R\vert$ (symmetric pushing force; SPHF in shorthand), due to $r_L=r_R$.
Next, in the PT broken symmetric phase, the characteristics of optical force have three situations: pushing, pushing, and null forces ( i.e., $\vec{P}_R=0$).
For the latter, we depict it by a gray dashing line.
Now, compared  our finding of this null force in Fig. 1 (c) with that in Fig. 1 (d), this null force belongs to bi-direction.
We call this bi-directional null force as BNF in shorthand.
Except for BNF, the optical force in the broken symmetric phase forms a flipping property, i.e., optical pulling from left (right) incidence but optical pushing from right (left) incidence.
In an exceptional point, the optical force can be pushing or one-way null force (OWNF), while the latter case is due to the absence of reflectionless for right illumination  i.e.,$r_R=0$ and $Im[M_{12}^{PT}]=0$.
We want to emphasize that in bi-reflectionless exceptional point, i.e., $r_R=r_L=0$, the optical force is BNF.

We remark that in unitary systems, due to power conservation and Lorentz reciprocity theorem, two reflectances are identical, $R_R=R_L$, irrespective of system configurations and any lossless materials embedded.
Thus the  optical forces can be pushing or null, and their magnitudes of forces by two individually opposite illuminations are equal.
Broking $R_R=R_L$ can be achieved by loss or gain embedded, resulting in non-Hermitian systems \cite{review1}.
In particular, one-way reflectionless is experimentally demonstrated in a staggered-layered structure made of silica and Si, associated with the emergence of an exceptional point \cite{ep1}.
Moreover, unbalanced gain and loss in spatial distribution can also have CPAL, already demonstrated in Ref. \cite{nonpt1}.

 \begin{figure*}[t]
\centering
\includegraphics[width=18cm]{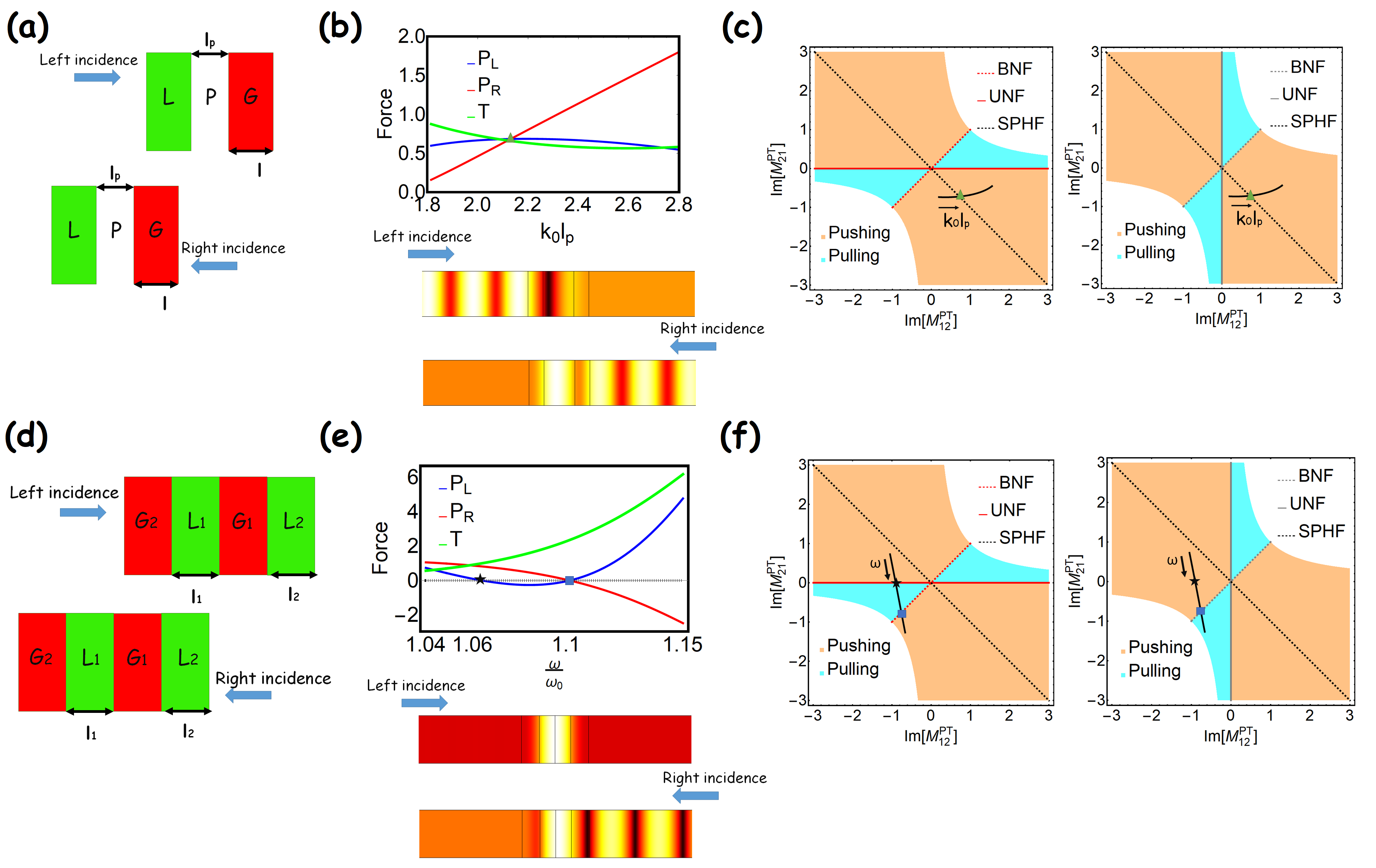}
\caption{(Color online) (a) Schematic of our studying PT symmetric system, made of a gain (G), vacuum (P), and loss (L) slabs. The thickness of gain/loss slabs is $l$, while the vacuum distance is $l_p$. We calculate the corresponding transmittance $T$, optical forces $P_L$ and $P_R$, by scanning the vacuum thickness $l_p$ in  the upper panel of (b), depicted by the green, blue, and red lines. All optical forces are pushing. The green triangle denotes $P_L=P_R$, i.e., SPHF. In the bottom panel, it shows the intensity of electric fields by two individual incidences from left and right for this SPHF (green triangle case). In order to see the relationship between the optical force and the PT phases for the (a) system, in (c), we study this system by means of the generalized parametric space. In the left and right panels of (c), they are from left and right illumination, respectively. This system under all scanning $l_p$ is operated at the symmetric phase with $T<1$. Except for PSR, the corresponding optical forces are pushing and unequal.
 The black star denotes SPHF.
We also demonstrate another PT-symmetric system made of two different constitution units of gain-loss slabs, shown in (d). Here the thickness for two gain-loss units are different, with $l_1$ and $l_2$. In (e), same as (b), but for the corresponding (d) system.
The black star and blue square represent UNF and BNF, respectively.
In the bottom panel of (e), we show  the intensity of electric fields for  UNF (the black star case).
In (f), same as (c), but for the (d) system. 
Now, it clearly demonstrates that the system for the black star and blue square shown in (e) are operated at the exceptional point and broken symmetric phase, respectively. } 
\end{figure*}  

\subsection{Material dispersion relation of permittivity for gain and loss}
In the following analysis, we adopt a Lorentz form of permittivity for gain
\begin{equation}
\epsilon_g(\omega)=\epsilon_0-\frac{\omega_p^2}{\omega_{r}^2-\omega^2-i\omega\gamma}
\end{equation}
where $\omega$ is angular operating  frequency, $\omega_w$ is angular  plasma  frequency, $\omega_r$ is angular resonance  frequency, and $\gamma$ is amplification linewidth .
This material dispersion is satisfied by Causality and Kramers-Krongie relations \cite{lorentz}.
For loss permittivity, we use $\epsilon_l(\omega)=\epsilon_g^{*}(\omega)$.
We set $\epsilon_0=1$ in the following calculation.

\subsection{Examples:  optical force under a single plane wave incidence}
Here we provide two examples with different PT symmetric system configurations, material parameters, and operating frequency, to verify our findings.
 We first consider a simple PT symmetric structure constituted of a gain, vacuum, and a loss slabs, as shown in Fig. 2 (a).
Here the thickness of gain/loss slab is $l$ and the vacuum thickness is $l_p$.
Now, we use the following parameters for this PT-symmetric system with $k_0l=1.036$, $\omega=0.495\omega_0$, $\omega_r=\omega_0$, $\omega_p^2=1.1\omega_0^2$, and $\gamma=1.3\omega_0$.
Here $\omega_0$ is angular reference frequency.
We tune the vacuum thickness $k_0l_p$ from $1.8$ to $2.8$.
In the upper panel of Fig. 2 (b), we show the corresponding transmittance $T$ and the optical forces $P_R$ and $P_L$ in a green, red, and blue lines, respectively.
In our scanning $k_0l_p$, all transmittances are lower than unity and the optical forces are unequal and positive, i.e., asymmetric pushing force. 
We see $P_L>P_R$ before $k_0l_p<2.13$, after that it becomes $P_L<P_R$.
In $k_0l_p=2.13$, the magnitudes of two optical forces is  equal $P_L=P_R$ (SPHF), marked by the green triangle.
Then, by the COMSOL Multiphysics,  in the bottom panel of Fig. 2 (b), we provide the intensity of electric fields for the SPHF case (green triangle).
The field distribution has the parity symmetry by two individual opposite illuminations, which supports this SPHF.
In order to further understand the behind PT-symmetric scattering responses and behaviors, we analyze  this PT-symmetric system by means of the parametric space in Fig 2 (c).
Here the black solid line represents this system by sweeping $l_p$.
It clearly displays that the system within this scanning $l_p$ operates at the symmetric phase, so that it guarantees to have $T<1$ and asymmetric (except for PSR) pushing forces.
Moreover, when $k_0l_p=2.13$, the system is at PSR, so $P_L=P_R$ (SPHF).

Then, we discuss another example with a more complicated system configuration made of two different gain-loss units, in Fig. 2 (d): in $G_1$, we set $\omega_{p1}^2=1.3\omega_0^2$, $\omega_{r1}^2=1.1\omega_0^2$, and $\gamma_1=1.3\omega_0$, while in $G_2$, we set $\omega_{p2}^2=4.3\omega_0^2$, $\omega_{r2}^2=\omega_0^2$, and $\gamma_{2}=1.2\omega_0$.
The geometry parameters for this system are $k_0l_1=1.036$ and $k_0l_2=1.175$.
Here we tune the operating frequency from $1.04\omega_0$ to $1.15\omega_0$.
We calculate the transmittance $T$, $P_R$ and $P_L$ in the upper panel of Fig. 2 (e).
Within $\omega=[1.04\omega_0,1.06\omega_0]$, we see $T<1$ and $P_R>P_L$.
However, when $\omega=1.06\omega_0$, we see $T=1$, $P_L=0$ and $P_R>0$, belonging to an exceptional point, marked by the black star.
Then $\omega=[1.06\omega_0,1.1\omega_0]$, we see $T>1$  $P_L<0$, and $P_R>0$.
In this situation, the optical force by left incidence becomes pulling, while in right incidence the optical force is pushing.
This is a flipping phenomenon for the optical force, depending on the incident directions.
Next,  when $\omega=1.1\omega_0$, it is a  BNF, i.e., $P_L=P_R=0$, also with $T>1$, marked by the blue square.
Next, $\omega>1.1\omega_0$, we again observe a pushing-pulling flipped force, while $P_L>0$ and $P_R<0$, also with $T>1$.
In the bottom panel of Fig. 2 (e), we provide the intensity of electric field for the corresponding case of $P_L=0$ and $P_R>0$ (the black star), implemented by the COMSOL Multiphysics.
The field distribution supports this exceptional point result due to uni-directional reflectionless.
In order to investigate the PT-symmetric scattering responses and behaviors, we use the generalized parametric space to study this system shown in Fig. 2 (f).
In the beginning, $\omega=[1.04\omega_0,1.06\omega_0]$, the system is at the symmetric phase, so that $T<1$ and there are unequal pushing forces.
However, when $\omega=1.06\omega_0$, it meets the exceptional point, so that $P_L=0$ and $P_R>0$, marked by the black star.
Then $\omega=[1.06\omega_0,1.1\omega_0]$, the system comes to the PT broken symmetric phase with $T>1$.
It is interesting to see that there shows the pulling-pushing flipped force.
Next, when $\omega=1.1\omega_0$, it has the BNF, $P_L=P_R=0$, marked by the blue square.
Then $\omega>1.1\omega_0$, we can see $P_L>0$, $P_R<0$, and $T>1$, belonging to the broken symmetric phase.
We stress that by means of the parametric space, the directionality and magnitudes of the optical forces can be clearly indicated, independent of system configurations, choices of material parameters, and operating frequency.

 \begin{figure*}
\centering
\includegraphics[width=18cm]{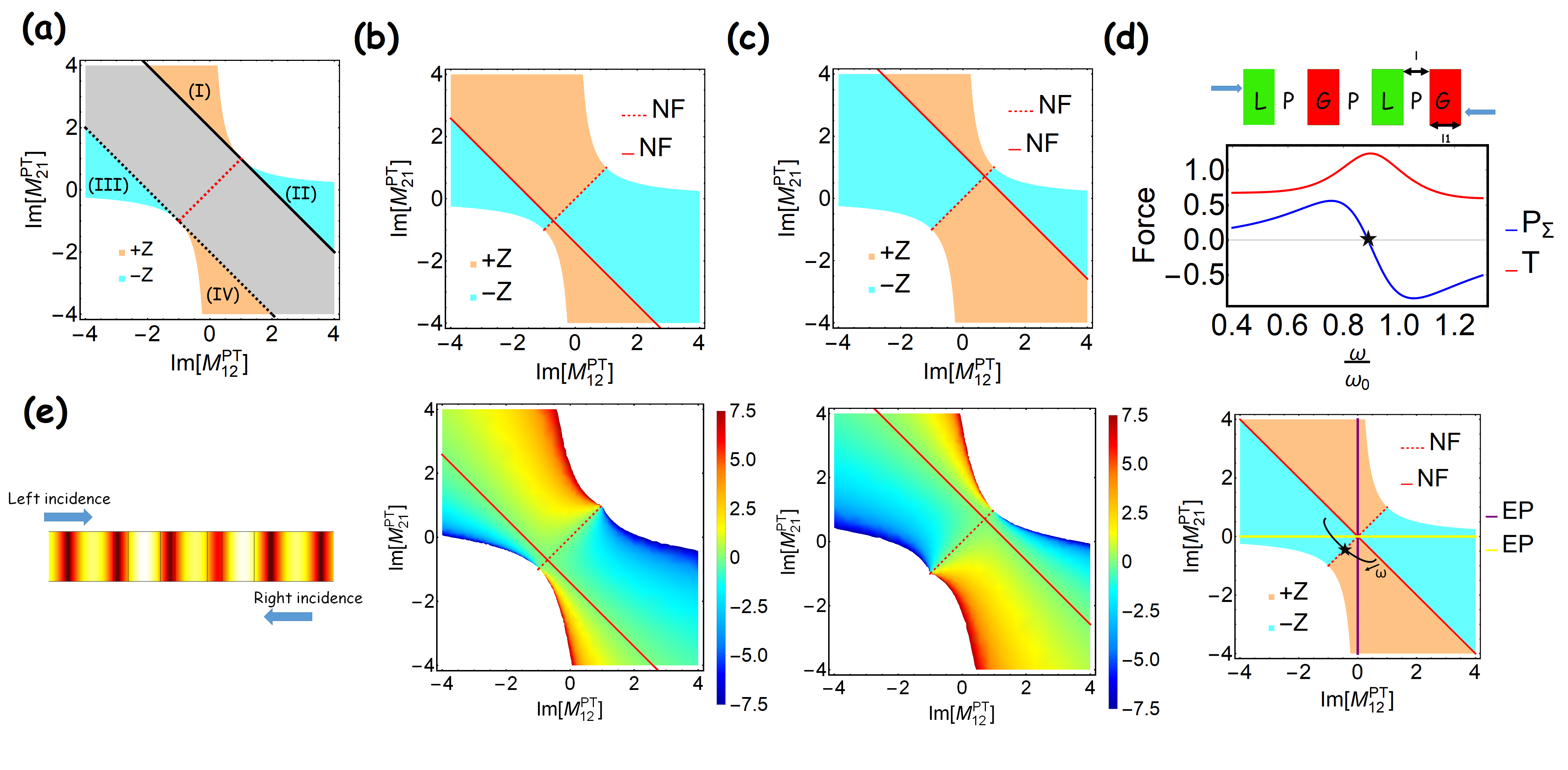}
\caption{(Color online) Under two counter-propagating plane waves excitation,  we show the region plot to indicate the directionality of $\vec{P}_{\sum}$ in (a). Here the water-blue regions denote the resultant force toward $-Z$ direction, while the dark-orange regions denote that toward $+Z$ direction. The gray region means undetermined directionality for the optical force, because it can be changed by $\sigma$. The solid and dashing black lines correspond to the upper and bottom boundary of undetermined one.  The red dashing line corresponds to the null force, independent of $\sigma$. When we choose $\sigma=-0.25\pi$, the directionality and magnitudes of the optical forces are shown in (b). The red solid line is also null force, but can be tuned by $\sigma$.  In (c) , same as (b), but we choose $\sigma=0.25\pi$. We design a PT-symmetric system made of double layers with identical gain-vacuum-loss slabs units shown in the upper panel of (d). Here We choose the relative phase $\sigma=0$. In the middle panel, it shows the corresponding transmittance $T$ and resultant optical force $P_{\sum}$ under the scanning frequency $\omega$ from $0.4\omega_0$ to $1.3\omega_0$. The black star means the null force. In the bottom panel, we study this system by means of the generalized parametric space, depicted by the black line. 
To further confirm the $P_{\sum}=0$,  we thus provide the intensity of electric field by the COMSOL Multiphysics  in (e), with expectation of same field distribution in the left and right leads.
} 
\end{figure*}

\section{Optical force by two counter-propagating plane waves incidence}
In this section, we discuss a more complicated situation where  PT symmetric systems encounter two counter-propagating plane waves impinging. 
Two plane waves have same optical intensity.
Without loss any generality, we let $E_L^{+}=E_0$ and $E_R^{-}=E_0e^{i\sigma}$ where $\sigma$ is a tunable relative phase with definition of $\sigma=[-\pi,\pi]$. 
With Eqs.(\ref{parametric}) and (\ref{two}), the corresponding optical force is
\begin{widetext}
\begin{equation}\label{twoforce}
\vec{P}_{\sum}=\frac{1}{2}\epsilon_0E_0^2\frac{Im[M_{21}^{PT}]-Im[M_{12}^{PT}]}{1-Im[M_{12}^{PT}]Im[M_{21}^{PT}]}(Im[M_{12}^{PT}]+Im[M_{21}^{PT}]-2\sin\sigma)\hat{z}.
\end{equation} 
\end{widetext}
The resultant optical force involves $\sigma$.
In Fig. 3 (a), we show the directionality of optical force, where the dark-orange regions denote the optical force toward $+Z$ direction, the water-blue regions denote that toward $-Z$ direction, and the gray color region can be either one depending on $\sigma$.
In addition, there have the bounds for this undetermined region indicated by the black solid and dashing curves, corresponding to $Im[M_{21}^{PT}]=-Im[M_{12}^{PT}]+2$ and $Im[M_{21}^{PT}]=-Im[M_{12}^{PT}]-2$, when $\sigma=0.5\pi$ and $\sigma=-0.5\pi$, respectively, in Fig. 3 (a).
Interestingly, by any values $\sigma$, some regions can support homogeneous property of the optical force, marked by (I)-(IV).
In the (I) and (IV) regions, although $\sigma$ can change the magnitude of $P_{\sum}$, its directionality of the optical force is always toward $+Z$ direction.
In the (II) and (III) regions, the optical force is always toward $-Z$ direction \cite{note1}.
We also find that PT-symmetric system can support the null force, when it meets $Im[M_{21}^{PT}]=Im[M_{12}^{PT}]$ or $Im[M_{12}^{PT}]+Im[M_{21}^{PT}]=2\sin\sigma$, obtained from Eq. (\ref{twoforce}).
For the former, this null force is independent of any $\sigma$, marked by a red dashing curve in Fig. 3 (a), located in the PT broken symmetry and exceptional point.
For the latter, obviously, this null force depends on $\sigma$.
The boundary of null optical force in the generalized parametric space would be restricted within $Im[M_{21}^{PT}]=-Im[M_{12}^{PT}]+2$ and $Im[M_{21}^{PT}]=-Im[M_{12}^{PT}]-2$, marked by black solid and dashing curves of Fig. 3 (a).

In Figs. 3 (b) and  (c), we provide the region and contour plots for the directionality and magnitudes of resultant optical force by using $\sigma=-0.25\pi$ and $\sigma=0.25\pi$, respectively.
We can observe that except for the fixed null force corresponding to the red dashing curve, with different relative phase $\sigma$, the optical force would be changed in magnitudes and directionality.

\subsection{Example:  optical force under two counter-propagating plane waves incidence}
To verify our findings,  we design a PT-symmetric system constituted of double layers with the identical unit of gain-vacuum-loss slabs, shown in the upper panel of Fig. 3 (d).
Here the parameters we use are $\omega_p^2=1.3\omega_0^2$, $\omega_r^2=1.1\omega_0^2$, $\gamma=\omega_0$, $k_0l_1=0.922$, and $k_0l=2.038$.
We choose the relative phase $\sigma=0$.
In the middle of Fig. 3 (d), we show the transmittance $T$ and magnitude of the optical force $P_{\sum}$.
When $0.4\omega_0<\omega<0.88\omega_0$, the optical force is toward $+Z$ direction, $P_{\sum}>0$.
At $\omega=0.88\omega_0$, it is null force $P_{\sum}=0$, marked by the black star.
Then $0.88\omega_0<\omega<1.3\omega_0$, the force is toward $-Z$, $P_{\sum}<0$.
For the transmittance $T$, we can see that  $T<1$ when $0.4\omega<\omega<0.79\omega_0$ and $\omega_0<\omega<1.3\omega_0$, $T=1$ when $\omega=0.79\omega_0$ and $\omega=\omega_0$, $T>1$ when $0.79\omega_0<\omega<\omega_0$.
For the case of $P_{\sum}=0$, we study its intensity of the electric field by the COMSOL Multiphysics as shown in Fig. 3 (e).
We see that the electric fields in the left and right leads are identical, supporting this $P_{\sum}=0$.
Again, in order to see the underlying PT scattering phases and responses, we study this system by means of the generalized parametric space in the bottom panel of Fig. 3 (d).
In the beginning $0.4\omega<\omega<0.79\omega_0$, the system is at symmetric phase and has the resultant force toward $+Z$ direction.
Then at $\omega=0.79\omega_0$, it is exceptional point, so $T=1$, and optical force is toward $+Z$ direction.
Then at $0.79\omega_0<\omega<\omega_0$, this system becomes PT broken symmetric phase, but its optical force is still toward $+Z$ direction.
At $\omega=0.88\omega_0$, corresponding to the black star, its force becomes null.
Then at $0.88\omega_0<\omega<\omega_0$, it is the broken symmetric phase, but the optical force becomes toward $-Z$ direction.
At $\omega=\omega_0$, it meets the exceptional point, while the optical force is toward $-Z$ direction.
After $\omega>\omega_0$, the optical force is still toward $-Z$ direction, but the system belongs to the PT symmetric phase.

\section{Conclusion}
We have discussed a generalized relation between the optical force and PT phases under a single and two counter-propagating plane waves illumination.
The generalized parametric space is deduced from the consideration of intrinsic PT symmetry condition and Lorentz reciprocity theorem applied to transfer matrix.
This parametric space can reflect any PT phases, extraordinary waves, and optical forces.
Under a single plane wave excitation, except of PSR, the optical forces are pushing and unequal in PT symmetric phase.
In PSR, the optical forces are pushing and have same magnitudes.
In the exceptional point, it can be pushing forces, UNF, or BNF (due to bi-directional reflectionless).
In the broken symmetric phase, the optical force can be pushing, pulling, or BNF. 
The optical pushing or pulling force has a flipped property: pulling from one incidence, but pushing from another incidence.
Under simultaneous illumination of two counter-propagating plane waves, the directionality and magnitudes of resultant optical force would be related to the relative phase $\sigma$.
However, there has a null force, regardless of  $\sigma$, in the specific regime of broken symmetric phase and exceptional point.
Several PT symmetric systems are demonstrated to verify these findings.
We envision that this work would have an impact on PT-symmetric optomechanics.

\section*{Acknowledgements}

This work was supported by Ministry of Science and Technology, Taiwan (MOST) ($107$-
$2112$-M-$143$-$001$-MY3).

 \section*{Appendix: Derivation of optical pressure under two counter-propagating plane waves illumination}
 We derive the formula for the optical force by two counter-propagating plane waves normal incident to an infinitely extended planar interface, as shown in Fig. 1 (a).
 
 The Maxwell's stress tensor represented in Cartesian coordinate is 
 \begin{widetext}
 \begin{equation}\label{tensor}
\begin{split}
\hat{T}&=\epsilon_0\vec{E}\otimes\vec{E}+\mu_0\vec{H}\otimes\vec{H}-\frac{1}{2}(\epsilon_0E^2+\mu_0H^2)\hat{I}\\
&=\begin{bmatrix}
\epsilon_0(E_x^2-\frac{E^2}{2})+\mu_0(H_x^2-\frac{H^2}{2}) & \epsilon_0E_xE_y+\mu_0H_xH_y & \epsilon_0E_xE_z+\mu_0H_xH_z\\
\epsilon_0E_xE_y+\mu_0H_xH_y & \epsilon_0(E_y^2-\frac{E^2}{2})+\mu_0(H_y^2-\frac{H^2}{2}) & \epsilon_0E_yE_z+\mu_0H_yH_z\\
\epsilon_0E_xE_z+\mu_0H_xH_z & \epsilon_0E_yE_z+\mu_0H_yH_z & \epsilon_0(E_z^2-\frac{E^2}{2})+\mu_0(H_z^2-\frac{H^2}{2}).
\end{bmatrix}
\end{split}
\end{equation}
 \end{widetext}
where $\vec{E}$ and $\vec{H}$ are real electric and magnetic fields, $\otimes$ is dyadic product, and $\hat{I}$ is second-order unit tensor \cite{book2}.

The total real electric and magnetic fields in the left lead are $\vec{E}_L=Re[E_L^{+}e^{i(k_0z-\omega t)}]\hat{x}+Re[E_L^{-}e^{i(-k_0z-\omega t)}]\hat{x}$ and $\vec{H}_L=\sqrt{\frac{\epsilon_0}{\mu_0}}Re[E_L^{+}e^{i(k_0z-\omega t)}]\hat{y}-\sqrt{\frac{\epsilon_0}{\mu_0}}Re[E_L^{-}e^{i(-k_0z-\omega t)}]\hat{y}$, respectively.
By Eq.(\ref{tensor}), the optical stress exerted on the left planar interface is
\begin{equation}
\begin{split}
\hat{T}\cdot (-\hat{z})&=\frac{\epsilon_0}{2}E_x^2\hat{z}+\frac{\mu_0}{2}H_y^2\hat{z}\\
&=\frac{\epsilon_0}{2}(Re[E_L^{+}e^{i(k_0z-\omega t)}]+Re[E_L^{-}e^{i(-k_0z-\omega t)}])^2\hat{z}\\
&+\frac{\epsilon_0}{2}(Re[E_L^{+}e^{i(k_0z-\omega t)}]-Re[E_L^{-}e^{i(-k_0z-\omega t)}])^2\hat{z}
\end{split}
\end{equation}
where the unit normal vector of the left planar surface is $-\hat{z}$.
Then the time-average optical pressure (the optical force per unit area) is
\begin{equation}
\vec{P}_L=<\hat{T}\cdot (-\hat{z})>_T=\frac{\epsilon_0}{2}(\vert E_L^{+}\vert^2+\vert E_L^{-}\vert^2)\hat{z}
\end{equation}
For the right lead, the total real electric and magnetic fields are $\vec{E}_R=Re[E_R^{+}e^{i(k_0z-\omega t)}]\hat{x}+Re[E_R^{-}e^{i(-k_0z-\omega t)}]\hat{x}$ and $\vec{H}_R=\sqrt{\frac{\epsilon_0}{\mu_0}}Re[E_R^{+}e^{i(k_0z-\omega t)}]\hat{y}-\sqrt{\frac{\epsilon_0}{\mu_0}}Re[E_R^{-}e^{i(-k_0z-\omega t)}]\hat{y}$, respectively.
By the same calculation steps, we obtain the time-average optical pressure exerted on the right planar interface,
\begin{equation}
\vec{P}_R=<\hat{T}\cdot (+\hat{z})>_T=-\frac{\epsilon_0}{2}(\vert E_R^{+}\vert^2+\vert E_R^{-}\vert^2)\hat{z}
\end{equation}
here the unit normal vector for the right planar surface is $+\hat{z}$. 

The total  optical pressure under two counter-propagating plane waves $\vec{P}_{\sum}$ is 
\begin{equation}\label{two}
\vec{P}_{\sum}=\vec{P}_L+\vec{P}_R=\frac{\epsilon_0}{2}(\vert E_L^{+}\vert^2+\vert E_L^{-}\vert^2-\vert E_R^{+}\vert^2-\vert E_R^{-}\vert^2)\hat{z}.
\end{equation}
Alternatively, there has a direct and intuitive way to see this formula by considering conservation of linear momentum of photons: 
in the left lead, incoming $E_L^{+}$ and outgoing $E_L^{-}$ waves would lead to an optical pressure toward $+z$ direction, while in the right lead, $E_R^{+}$ and $E_R^{-}$ waves would lead to that toward $-z$ direction.

Next, by the scattering matrix formula Eq.(\ref{scattering}), Eq.(\ref{two}) becomes
 \begin{widetext} 
\begin{equation}
\begin{split}
P_{\sum}\hat{z}=\frac{1}{2}\epsilon_0\{\vert E_L^{+} \vert^2 (1-T+R_L)-\vert  E_R^{-} \vert^2(1-T+R_R)+2Re[ E_L^{+}  E_R^{-*}r_Lt^{*} ]-2Re[E_L^{+} E_R^{-*}r_R^{*}t]\}\hat{z}.
\end{split}
\end{equation}
\end{widetext}


\begin{thebibliography}{99}

\bibitem{book1}
M. Mansuripur, \textit{Field, Force, Energy and Momentum in Classical
Electrodynamics} (Bentham Science Publishers, United Arab
Emirates, 2011).

\bibitem{book2}
L. Novotny and Bert Hecht, \textit{ Principles of nano-optics} (Cambridge university press, 2012)

\bibitem{nobel1}
A. Ashkin and J. M. Dziedzic,``Observation of Resonances in the Radiation Pressure on Dielectric Spheres,''
Phy. Rev. Lett.  \textbf{38}, 1351 (1977).


\bibitem{particle1}
A. Rahimzadegan, R. Alaee, I. Fernandez-Corbaton, and C. Rockstuhl,
``Fundamental limits of optical force and torque,''
Phys. Rev. B \textbf{95}, 035106 (2017).

\bibitem{beam1}
J. Chen, J. Ng, Z. Lin, and C. T. Chan,
``Optical pulling force,''
Nat. Photonics \textbf{5}, 531-534 (2011).



\bibitem{beam2}
H. Chen, S. Liu, J. Zi, and Z. Lin,
``Fano Resonance-Induced Negative Optical Scattering Force on Plasmonic Nanoparticles,''
ACS Nano \textbf{9}, 1926 (2015).

\bibitem{beam3}
J. J. Saenz,
``Laser tractor beams,'' Nat. Photonics \textbf{5}, 514 (2011).



\bibitem{plane1}
R. Ali, F. A. Pinheiro, R. S. Dutra, and P. A. Maia Neto,
``Tailoring optical pulling forces with composite microspheres,''
Phys. Rev. A \textbf{102}, 023514 (2020).


\bibitem{plane2}
E. Mobini, A. Rahimzadegan, C. Rockstuhl, and R. Alaee,
``Theory of optical forces on small particles by multiple plane waves,''
J. Appl. Phys. \textbf{124}, 173102 (2018).

\bibitem{plane3}
A. Y. Bekshaev, K. Y. Bliokh, and F. Nori,
``Transverse Spin and Momentum in Two-Wave Interference,''
Phys. Rev. X \textbf{5}, 011039 (2015).



\bibitem{beam5}
S.-H. Lee, Y. Roichman, and D. G. Grier,
``Optical solenoid beams,''
Opt. Express \textbf{18}, 6988-6993 (2010). 

\bibitem{beam6}
S. Lepeshov and A. Krasnok,
``Virtual optical pulling force,''
 	Optica \textbf{7}, 1024-1030 (2020).


\bibitem{gain1}
A. Mizrahi and Y. Fainman, ``Negative radiation pressure on gain medium structures,''
Opt. Lett. \textbf{35}, 3405-3407 (2010).

\bibitem{gain2}
K. J. Webb and Shivanand, ``Negative electromagnetic
plane-wave force in gain media,'' Phys. Rev. E \textbf{84}, 057602
(2011).

\bibitem{gain3}
D. Gao, R. Shi, Y. Huang, and L. Gao,
``Fano-enhanced pulling and pushing optical force on active plasmonic nanoparticles,''
Phys. Rev. A \textbf{96}, 043826 (2017).


\bibitem{pt1}
S. K. Özdemir, S. Rotter, F. Nori, and L. Yang,
``Parity–time symmetry and exceptional points in photonics,'' Nat. Mater. \textbf{18}, 783 (2019).

 
 \bibitem{pt2}
 Y. D. Chong, L. Ge, and A. D. Stone, ``PT-Symmetry Breaking and Laser-Absorber Modes in Optical Scattering Systems,'' Phys. Rev. Lett. \textbf{106}, 093902 (2011).
 
 \bibitem{pt3}
 L. Ge, Y. D. Chong, and A. D. Stone, ``Conservation relations and anisotropic transmission resonances in one-dimensional
PT -symmetric photonic heterostructures,'' Phys. Rev. A \textbf{85}, 023802 (2012).


 \bibitem{pt4}
 S. Longhi, ``PT -symmetric laser absorber," Phys. Rev.
A \textbf{82}, 031801(R) (2010).

\bibitem{pt5}
Z. Lin, H. Ramezani, T. Eichelkraut, T. Kottos, H. Cao,
and D. N. Christodoulides, ``Unidirectional Invisibility
Induced by PT-Symmetric Periodic Structures," Phys.
Rev. Lett. \textbf{106}, 213901 (2011).

\bibitem{pt6}
K. G. Makris, R. El-Ganainy, D. N. Christodoulides, and
and Z. H. Musslimani, ``Beam Dynamics in PT Symmetric
Optical Lattices," Phys. Rev. Lett. \textbf{100}, 103904
(2008).

\bibitem{pt7}
S. Longhi, ``Bloch Oscillations in Complex Crystals with
PT Symmetry," Phys. Rev. Lett. \textbf{103}, 123601 (2009).

\bibitem{pt8}
R. Fleury, D. L. Sounas, and A. Alu, ``Negative Refraction and Planar Focusing Based on Parity-Time Symmetric Metasurfaces," Phys. Rev. Lett. \textbf{113}, 023903 (2014).

\bibitem{pt9}
M. Farhat, M. Yang, Z. Ye, and P.-Y. Chen, ``PT-symmetric
absorber-laser enables electromagnetic sensors with unprecedented sensitivity,'' ACS Photonics \textbf{7}, 2080 (2020).

\bibitem{pt10}
 P.-Y. Chen and J. Jung, ``PT Symmetry and Singularity-Enhanced
Sensing Based on Photoexcited Graphene Metasurfaces,'' Phys.
Rev. Appl. \textbf{5}, 064018 (2016).

\bibitem{pt11}
L. Chang, X. Jiang, S. Hua, C. Yang, J. Wen, L. Jiang, G. Li,
G. Wang, and M. Xiao, ``Parity-time symmetry and variable optical isolation in active–passive-coupled microresonators,'' Nat.
Photonics \textbf{8}, 524 (2014).


\bibitem{pt12}
H. Ramezani, T. Kottos, V. Kovanis, and D. N. Christodoulides, ``Exceptional-point
dynamics in photonic honeycomb lattices with PT symmetry,'' Phys. Rev. A \textbf{85}, 013818 (2012).

 
 
 \bibitem{ptforce1}
 R. Alaee, J. Christensen, and M. Kadic,
 ``Optical Pulling and Pushing Forces in Bilayer PT -Symmetric Structures,''
 Phys. Rev. Applied \textbf{9}, 014007 (2018).
 
  \bibitem{ptforce2}
  R. Alaee, B. Gurlek, J. Christensen, and M. Kadic,
 ``Optical force rectifiers based on PT-symmetric metasurfaces,''
 Phys. Rev. B \textbf{97}, 195420 (2018).
 
 
 

\bibitem{pt13}
J. Y. Lee and P.-Y. Chen,
``Generalized parametric space, parity symmetry of reflection, and systematic design approach for parity-time-symmetric photonic systems,''
Phys. Rev. A (to be appeared).
 
 
\bibitem{review1}
R. El-Ganainy, K. G. Makris, M. Khajavikhan, Z. H.
Musslimani, S. Rotter, and D. N. Christodoulides,
``Non-Hermitian Physics and PT Symmetry,'' Nat. Phys. \textbf{14}, 11 (2018).


\bibitem{ep1} 
 L. Feng, X. F. Zhu, S. Yang, H. Y. Zhu, P. Zhang, X. B. Yin,
Y. Wang, and X. Zhang, ``Demonstration of a Large-Scale
Optical Exceptional Point Structure,'' Opt. Express \textbf{22}, 1760
(2014).
 
 
 \bibitem{nonpt1}
 M. Sakhdari, N. M. Estakhri, H. Bagci, and P.-Y. Chen,
``Low-Threshold Lasing and Coherent Perfect Absorption in Generalized
PT -Symmetric Optical Structures,''
Phys. Rev. Applied \textbf{10}, 024030 (2018).

\bibitem{lorentz}
A. A. Zyablovsky, A. P. Vinogradov, A. V. Dorofeenko,
A. A. Pukhov, and A. A. Lisyansky, ``Causality and phase
transitions in PT-symmetric optical systems," Phys. Rev.
A \textbf{89}, 033808 (2014).




\bibitem{note1}
This result can be obtained by analyzing two conditions: $Im[M_{21}^{PT}]-Im[M_{12}^{PT}]$ and $Im[M_{12}^{PT}]+Im[M_{21}^{PT}]-2\sin\sigma$.
In the (I) region, it can guarantee to have $Im[M_{21}^{PT}]>Im[M_{12}^{PT}]$ and $Im[M_{12}^{PT}]+Im[M_{21}^{PT}]-2\sin\sigma>0$. Overall, the resultant force in this region can have $P_{\sum}>0$.
For other regions (II)-(IV), we can use the same approach to indicate its directionality of the resultant force.

 
\end{thebibliography}
\end{document}